# Using top graphene layer as sacrificial protection during dielectric atomic layer deposition

Xiaohui Tang[1,*], Nicolas Reckinger[2,3], Olivier Poncelet[1], Pierre Louette[3], Jean-François Colomer[2,3], Jean-Pierre Raskin[1], Benoit Hackens[4], and Laurent A. Francis[1]

[1]ICTEAM Institute, Université catholique de Louvain, Place du Levant 3, 1348 Louvain-la-Neuve, Belgium.

[2]Research Group on Carbon Nanostructures (CARBONNAGe), University of Namur, Rue de Bruxelles 61, 5000 Namur, Belgium.

[3]Department of Physics, Research Center in Physics of Matter and Radiation (PMR), University of Namur, 61 Rue de Bruxelles 61, 5000 Namur, Belgium.

[4] NAPS/IMCN, Université catholique de Louvain, 2 Chemin du Cyclotron, 1348 Louvain-la-Neuve, Belgium

E-mail: xiaohui.tang@uclouvain.be
Phone: +32 (0)10472589. Fax: +32 (0)10472598

**Abstract**

We investigate the structural damage of graphene underlying dielectrics ($HfO_2$ and $Al_2O_3$) by remote plasma-enhanced atomic layer deposition (PE-ALD). Dielectric film is grown on bilayer graphene without inducing significant damage to the bottom graphene layer. Based on Raman spectra, we demonstrate that the bottom graphene layer has the salient features of single layer graphene. During the initial half-cycle PE-ALD, the upper graphene layer reacts with the metal precursor, forming uniform nucleation islands or an active metallic carbide layer. After monolayer dielectric coverage, the bottom graphene layer has additional protection. The upper graphene layer serves as a sacrificial layer, which not only promotes the adhesion of dielectric on graphene, but also protects the lattice symmetry of the bottom graphene layer. Our results indicate that bilayer graphene allows for controlling/limiting the degree of defect during the ALD of dielectrics and could be a good starting material for building filed effect transistors and sensing devices.

## 1. Introduction

One of the most explored application domain for graphene is nanoelectronics because of its high carrier mobility and atomic thickness[1,2]. However, gate dielectric deposition is an important challenge for transferring graphene transistors from laboratory level to industrial production. Dielectric or metal deposition induces defects in single-layer (1L) graphene and at the interface between dielectric and multi-layer graphene[3]. The carrier mobility is very sensitive to the graphene lattice defects and interface quality. It has indeed been reported that the carrier mobility of suspended graphene is significantly higher than that of graphene lying on a silicon dioxide ($SiO_2$) substrate[4] due to corrugation, traps at the interface and fixed charges in the dielectric layer. Therefore, it is a crucial task to directly grow metal-oxide dielectric films on graphene without inducing damage into the graphene lattice.

Although physical vapor deposition (such as sputtering), widely used in semiconductor industry, can provide high deposition rates and preserve film stoichiometry, it generates extensive damage in graphene[5] from high energy sputtered atoms. Researchers mostly choose atomic layer deposition (ALD)[6] for dielectric growth on graphene[7]. ALD has a feature of high conformity and controls thickness and uniformity of deposited films to atomic-level precision while averting physical damage of energized atoms to the surface. ALD techniques are classified into plasma (oxygen-based) and thermal (water-based) depositions. Very few examples dealing with the former technique were reported. Nayfeh *et al.*[8] demonstrated a graphene transistor for which aluminum oxide ($Al_2O_3$) gate dielectric was directly deposited on graphene by using a remote plasma-enhanced ALD (PE-ALD) process. Most of the reports are related to the latter method, because plasma is rather aggressive (especially using a direct plasma) and generally leads to etching of graphene[9]. It is well known that graphene is hydrophobic and inert. Specifically, graphene does not provide reactive nucleation sites for the precursors in thermal ALD[10] since it does not display covalent bonds out of the plane. Therefore, growing high-quality and uniform-coverage dielectrics by thermal ALD requires a graphene pretreatment. Various approaches have been proposed: (i) graphene is chemically modified by fluorine[11], ozone[12], organic molecules[13], nitride oxide[14,15], or perylene tetracarboxylic acid[16]; (ii) metal particles are deposited on graphene as appropriate nucleation layers[17]; (iii) graphene islands, serving as a seed layer, are generated by low-power plasma[18]. Some of these approaches are complicated and incompatible with existing mainstream silicon technology. Particularly, these approaches might cause undesirable side effects, such as: leaving seed layers, creating defects, doping graphene, increasing dielectric thickness, and degrading dielectric properties. Two previous articles point to a possible degradation of graphene induced by the pretreatment[19,20]. Alternatively, the ozone pretreatment is proved to be responsible for significant damage to graphene in the high-temperature thermal ALD[21].

In this work, we use mild plasma conditions to directly grow hafnium oxide ($HfO_2$) and $Al_2O_3$ dielectrics on graphene by PE-ALD. In our process, the graphene samples are placed away from the plasma source for outside of the glow discharge. The remote oxygen plasma with low ion bombardment avoids fast etching of graphene while the reaction between graphene and the precursor still leads to physical/chemical modification of graphene. Based on Raman spectroscopy and X-ray photoelectron spectroscopy (XPS) analyses, we study the damage in 1L graphene underlying $HfO_2$ and $Al_2O_3$ dielectrics (hereafter referred to as $HfO_2$/graphene and $Al_2O_3$/graphene) upon different oxygen plasma power levels. We also investigate the level of damage for 1L, bi- (2L), and tri-layer (3L) graphene underlying $HfO_2$ and $Al_2O_3$ dielectrics for a fixed oxygen plasma power. Our results show that, in the case of PE-ALD $HfO_2$, the upper layer of 2L graphene serves as a sacrificial layer, which not only promotes the adhesion of $HfO_2$ with graphene, but also protects the lattice integrity of the bottom graphene layer. 2L graphene allows for controlling/limiting the defect formation during the PE-ALD $HfO_2$ process and could be a good starting material for certain applications, such as graphene-based transistors and sensing devices[22,23]. To date, wafer-scale homogeneous 2L graphene has been synthesized by chemical vapor deposition[24].

**2. Experiments**

2.1 Graphene chemical vapor deposition

Graphene is grown by atmospheric pressure chemical vapor deposition (APCVD) with methane as precursor on copper foils. Graphene is then transferred onto $SiO_2$/Si substrates by the usual method based on polymethyl methacrylate (PMMA)[25], after etching the copper foil in ammonium persulfate. More details can be found in reference[26]. 300-nm-thick $SiO_2$ is used to easily observe graphene with a conventional white light microscope. The layer number of the graphene flakes is identified by optical microscopy, scanning electron microscopy (SEM) and subsequently confirmed through Raman spectroscopy.

2.2 Dielectric atomic layer deposition

$HfO_2$ and $Al_2O_3$ films are deposited on graphene/$SiO_2$/Si stacks, by PE-ALD (Fiji F200 from Ultratech/Cambridge NanoTech Inc., MA) at 250 °C. The plasma source, inductively coupled at 13.56 MHz, is far away from the samples. The distance between the plasma source and sample location is larger than 40 cm. This is very important since the type and concentration of the reactive species, *i.e.* electrons, ions, radicals, strongly depend on this distance. Outside of the glow discharge, the excited species have a weak energy and are present in small number, only long lifetime species (radicals) are present while ions and electrons recombine quickly[27]. In order to remove the PMMA residues, the samples are heated up at 250 °C in vacuum.

During both dielectric film depositions, the pulse duration of oxygen plasma (oxidant precursor) is 10 s for each cycle. The metal and oxidant precursor pulses are separated by a short argon purge of 5 s. The other parameters related to the metal precursors and the final thickness of both dielectric films are listed in Table 1. The thickness of the dielectric films is measured by *in situ* ellipsometry from reference films directly deposited on Si substrates. The composition of the dielectric films is characterized by XPS.

Table 1: Process conditions in PE-ALD and thicknesses of the two dielectric films.

| Dielectric | $HfO_2$ | $Al_2O_3$ |
|---|---|---|
| Precursor acronym | TDMA-Hf | TMA |
| Precursor chemical formula | $[(CH_3)_2N]_4Hf$ | $Al_2(CH_3)_6$ |
| Precursor temperature (°C) | 75 | 25 |
| Precursor pulse duration (s) | 0.25 | 0.06 |
| Cycle number (cycle) | 55 | 55 |
| Final thickness (nm) | 7.9 | 5.5 |

In order to investigate the damage level of graphene upon different oxygen plasma power levels, $HfO_2$ and $Al_2O_3$ films are deposited on graphene/$SiO_2$/Si stacks with nominal 300 W and with reduced oxygen plasma power of 200 and 150 W, respectively.

2.3 Raman spectroscopy

The measurements are performed at room temperature by a LabRam Horiba spectrometer. The laser beam (wavelength of 514 nm) is focused on the center of hexagons and a 100× objective (NA = 0.95) is used to collect the signal. The incident power is kept below 1 mW. Low resolution (150 g/mm) and high resolution (1800 g/mm) gratings are used for the measurements.

2.4 XPS analyses

A ThermoFisher Scientific K-alpha spectrometer is employed. It is equipped with a monochromatized Al Kα1,2 x-ray source and a hemispherical deflector analyzer. The spectra are recorded at constant pass energy (150 eV for depth profiling and survey; 30 eV for high resolution spectra). A flood gun (low energy electrons and Ar ions) is used during all the measurements. During the sputtering, the Ar$^+$ ion gun is operated at a low energy (200 eV), with an erosion time of 5 s per cycle, and the analysis is done in snapshot mode. The approximate sputter rate (as calculated from the depth profiles) is 0.016 nm/s for HfO$_2$ and 0.022 nm/s for Al$_2$O$_3$, respectively. The XPS data are treated with the Avantage software. High resolution spectra are fitted by Gaussian-Lorentzian lineshapes with an Avantage "smart" background (*i.e.* a Shirley background in most cases, or a linear background in case the lineshape decreases with increasing BE).

## 3. Characterization

### 3.1 Scanning electron microscopy and optical microscopy observations

Figure 1 shows SEM images of APCVD graphene on copper foils. The different contrasts in the images correspond to the different graphene layer number. Form Fig. 1a-c, it can be seen that the 1L, 2L, and 3L graphene domains are hexagonal. Figure 2a illustrates an optical microscopy image of pristine graphene transferred to the SiO$_2$/Si substrate, revealing that a graphene film composed of isolated and contiguous hexagonal flakes of variable thickness. Most of the hexagons are 1L. 2L and 3L graphene hexagons are found in certain regions. As shown in Fig. 2b, graphene underlying the dielectric films is still clearly visible, indicating the uniform covering of graphene with the dielectric materials. No clustering or pinholes in the dielectric films is observed.

### 3.2 Graphene structural damage evaluation by Raman spectroscopy

#### 3.2.1 Structural damage of 1L graphene underlying different dielectrics

Raman spectroscopy is a nondestructive method and is employed to assess the structural damage of graphene underlying dielectric films. Figure 3 shows the Raman spectra of 1L graphene after PE-ALD Al$_2$O$_3$ and HfO$_2$, at an oxygen plasma power of 300 W. The spectra are offset for clarity. For the sake of comparison, the Raman spectrum of pristine 1L graphene is also shown in the figure. The peak at ~1588 cm$^{-1}$ originates from the G mode of graphene (a first-order Raman process). The non-perturbed G mode is usually around 1582 cm$^{-1}$ (see the work of Ferrari *et al.*[28]). The slight upshift is most probably due to residual strain originating from the copper substrate or/and unintentional doping. The doping possibly comes from traps in the SiO$_2$ substrate, from insufficient rinsing after copper etching, from PMMA residues in transfer step, from moisture in the air, and other similar contamination sources[29,30,31]. The other peak is the 2D mode at ~2687 cm$^{-1}$, which is a two-phonon second-order Raman process. The intensity ratio $I_{2D}/I_G$ is 1.9, and the full width at half maximum (FWHM) of the 2D peak is 31 cm$^{-1}$. These figures of merit confirm the presence of 1L graphene. For the Raman spectra of 1L graphene underlying HfO$_2$ and Al$_2$O$_3$, besides the peaks of pristine graphene, two peaks (defect-activated peaks) appear. Namely the D peak located at ~1356 cm$^{-1}$ and the D' peak (~1620 cm$^{-1}$) at the right shoulder of the G mode. The D mode in sp$^2$ graphene is a feature that is only observed when the crystal symmetry is broken by point defects or the edges of graphene[32]. The D' mode corresponds to an independent defect-assisted intravalley process in graphene. It could be due to the presence of sp$^3$ bonding. In pristine 1L graphene, the D peak is weak, as shown in the bottom spectrum. Since the size of the examined hexagon flake is large enough (larger than 15 µm from vertex to vertex) to make the measurement inside the crystalline region (in the center of hexagon) with a Raman laser spot diameter of about 1 µm, the boundary of the hexagons does not contribute to the spectrum here. After the dielectric depositions, the D peaks become very strong. This indicates that the dielectric depositions break the symmetry of the graphene lattice and induce structural defects in graphene. The positions of the G peaks are not shifted after dielectric depositions, the FWHM of 2D peaks are broadened and the D' peaks are separated from the G peaks. These characteristics indicate that graphene is disordered, but is still nanocrystalline[32]. We therefore use the intensity ratio ($I_D/I_G$) as a quantitative measurement of structural damage. Cançado *et al.*[33] proposed that the average distance between defects, $L_a$, for 1L graphene is:

$$L_a = (2.4 \times 10^{-10}) \lambda_{laser}^4 (I_D/I_G)^{-1} \text{ (nm)} \qquad (1)$$

With $\lambda = 514$ nm, $L_a$ is calculated to be 10 and 11 nm for the $HfO_2$/graphene and $Al_2O_3$/graphene, respectively. In contrast, for pristine 1L graphene, $L_a$ equals 118 nm.

3.2.2 Structural damage of 1L graphene upon different oxygen plasma power levels

In order to investigate the damage of graphene subjected to different oxygen plasma power levels, we decrease the oxygen plasma power from 300 to 200 W and 150 W for PE-ALD $HfO_2$ and PE-ALD $Al_2O_3$, respectively. The corresponding Raman spectra of 1L graphene underlying the two dielectrics are shown in Fig. 4a and 4b, respectively. The D and D' peaks exist even if the oxygen plasma power is reduced to 150 W. The intensity ratios of D peak to G peak ($I_D/I_G$) are 1.8 and 1.5 for $HfO_2$/graphene and $Al_2O_3$/graphene, respectively. The $I_D/I_G$ intensity ratios are very similar under the different power levels (300 and 200 W for $HfO_2$, 300 and 150 W for $Al_2O_3$), implying that the amount of generated disorders in graphene is not correlated to the plasma power levels in this range of powers. For ALD $Al_2O_3$ on graphene, Lim *et al.* used nitrogen plasma to pretreat graphene[34]. They investigated the dependence between the number of defects and the nitrogen plasma power levels (30, 60, and 100 W). Their results indicate that the number of defects is increased with the nitrogen plasma power level. Our results suggest that for oxygen plasma power above 150 W, the number of defects in graphene has reached saturation.

3.2.3 Structural damage of 2L and 3L graphene underlying the dielectrics

The layer number is first identified by the color contrast of graphene under optical microscopy, followed by Raman spectroscopy measurements. Figure 5a shows the Raman spectra of pristine graphene with various layer numbers. All the spectra are dominated by two main peaks corresponding the G and 2D modes. In the case of $Al_2O_3$/graphene (Fig. 5b), the amplitude of the D and D' peaks with respect to the G peak are attenuated with the number of layers, while the 2D peak positions are upshifted with the number of layers. In sharp contrast to $Al_2O_3$/graphene, the D peak intensity becomes very weak and the D' peak even almost disappears from the spectrum of 3L graphene for the case of $HfO_2$/graphene (Fig. 5c). For 2L graphene, the $I_{2D}/I_G$ intensity ratio is 1.88, and the FWHM of 2D peak is 34 cm$^{-1}$. Surprisingly, they are the same as the values of pristine 1L graphene (1.9 and 31 cm$^{-1}$ in table 2). Also, the D peak intensity is relatively low. Consequently, we can speculate that the upper layer of 2L graphene may serve as a sacrificial layer. It limits physical defects into the bottom graphene layer. Compared to the spectrum of pristine 1L graphene, the 2D peak of 2L graphene underlying $HfO_2$ is upshift. The operating temperature of PE-ALD process being 250 °C, most of the native stress from growth and transfer, as well as PMMA residues, are removed and it leads to a slight *p*-type doping induced by the excited species8[35,36]. It can also be seen that the spectrum line shape and $I_{2D}/I_G$ intensity ratio of 3L graphene are similar to that of pristine 2L graphene in Fig. 5a. This could be also attributed to the same protective role of the sacrificial top layer. Table 2 summarizes the positions of the G and 2D peaks, the $I_{2D}/I_G$ intensity ratio, and the 2D peak FWHM for pristine 1L and 2L graphene and for 2L and 3L graphene underlying $HfO_2$. All the data are obtained from measurements with high resolution (1800 g/mm) gratings.

Table 2: G and 2D peak positions, the $I_{2D}/I_G$ intensity ratio, and the 2D peak FWHM for pristine 1L and 2L graphene and for 2L and 3L graphene underlying $HfO_2$.

| Name | G peak position ($cm^{-1}$) | 2D peak position ($cm^{-1}$) | $I_{2D}/I_G$ intensity ratio | 2D peak FWHM ($cm^{-1}$) |
|---|---|---|---|---|
| Pristine 1L graphene | 1588 | 2687 | 1.9 | 31 |
| 2L graphene underlying $HfO_2$ | 1595 | 2697 | 1.88 | 34 |
| Pristine 2L graphene | 1586 | 2699 | 0.71 | 54 |
| 3L graphene underlying $HfO_2$ | 1588 | 2708 | 0.73 | 54 |

3.3 X-ray photoelectron spectroscopy measurements

We perform *ex situ* XPS measurements to evaluate the impact of PE-ALD dielectrics on graphene. Four samples are sputtered with an $Ar^+$ gun to perform a depth profile: 5.5-nm-thick $Al_2O_3$ and 7.9-nm-thick $HfO_2$ are deposited either on silicon (as references) or on graphene/$SiO_2$/silicon substrates (hereafter referred to as $Al_2O_3$/silicon, $HfO_2$/silicon, $Al_2O_3$/graphene, and $HfO_2$/graphene). Core level spectra are recorded from carbon (C 1*s*), oxygen (O 1*s*), hafnium (Hf 4*f*), aluminum (Al 2*p*), and silicon (Si 2*p*). The elemental composition of 20-nm-thick reference dielectric layers can be estimated from the ratios of the integrated intensities of the XPS spectra: [O/Hf] = 2.15±0.1 and [O/Al] = 1.47±0.05. These results testify to the good quality of the dielectrics. We now focus on the C 1*s* atomic concentration profile and spectra of each sample. Figures 6a and b illustrate the depth profiles of the $Al_2O_3$/silicon and $HfO_2$/silicon samples, respectively. In both cases, a small amount of carbon (2%) is found in the profiles (except for a ~15% concentration corresponding to adventitious carbon on top of the dielectric layers). Moreover, a slight increase of the carbon concentration is observed when approaching the interface between the dielectric and silicon, most likely originating from residual contamination on silicon before PE-ALD. Figures 6c and 6d display the depth profiles of the $Al_2O_3$/graphene and $HfO_2$/graphene samples, respectively. We can clearly identify the presence of graphene between the dielectric and the $SiO_2$/Si substrate. Figures 6e and f exhibit the C 1*s* spectra of the $Al_2O_3$/graphene and $HfO_2$/graphene samples at the maximum of the carbon profiles, respectively. The main peak at 284.4 eV in both spectra corresponds to graphene. Strikingly, in contrast to the $Al_2O_3$/graphene sample, the $HfO_2$/graphene sample displays an additional peak at 281.5 eV. This peak can be attributed to the formation of the metallic carbide Hf-C. Consequently, the $HfO_2$/graphene profile in Fig. 6d can be fitted by its two components: C in graphene and C in Hf-C. At the interface, the Hf-C concentration reaches 2% of the total composition. The question is: what is the origin of the formation of the chemical bond between Hf and C? It was reported by Engelhard[37] that $Ar^+$ sputtering of ALD $HfO_2$ induces the formation of Hf-C (at an ion-gun energy of 2 keV), amounting to only 1% of the total composition. However, since the ion gun is operated at the very low energy of 200 eV for our sputtering, it seems very unlikely that Hf-C is generated by the erosion process. Instead, we believe that Hf-C results from the reaction between Hf and graphene during the initial ALD cycle.

**4. Discussion**

The reason why the bottom layer of 2L graphene underlying PE-ALD $HfO_2$ presents pristine 1L-like features is discussed hereafter. During the initial PE-ALD half-cycle, the reaction between the TDMA-Hf precursor[38] and the upper graphene layer may create many islands on the top of the bottom graphene layer. A large amount of defects is concentrated in these islands, but the bottom graphene layer is preserved. Such defects serve as reactive nucleation sites for $HfO_2$ growth, similar to the nucleation treatments in thermal ALD as mentioned above. This hypothesis is supported by the work of Lupina et al..[39] They directly deposited $HfO_2$ film on 1L graphene in thermal ALD, benefiting from the 2L or multiple layer islands spontaneously originating from commercially available graphene films. Their results prove that not only graphene edges but also the 2L or multiple islands on the graphene surface provide natural nucleation sites for the growth of $HfO_2$. On the other hand, during the initial PE-ALD half-cycle, the upper graphene layer is gradually etched under the gentle plasma conditions. This results in the C-C bond scission between A/B or A/A stacking. The residual bonds may then react with the TDMA-Hf precursor during the second ALD half-cycle to form Hf-C, meanwhile the bottom graphene layer remains mostly unaffected (as proved by the corresponding Raman spectrum). Hf in Hf-C acts as a uniform and active layer for $HfO_2$ growth. The XPS data have demonstrated the presence of Hf-C bonds in Fig. 6 (f). However, in situ XPS investigations and high-resolution transmission electron microscopy analysis are needed to further elucidate the true mechanism and provide direct evidence for C islands or Hf-C bonds. Subsequently, the coverage of the first $HfO_2$ layer or Hf-C layer protects the bottom graphene layer from damage by the following deposition cycles. As seen in the Raman spectra of $Al_2O_3$/graphene (in Fig. 5b), 1L, 2L, and 3L graphene are significantly damaged in the PE-ALD process. This is due to the fact that the TMA precursor by itself does not react with graphene at low temperature (lower than 300 °C)[40] and therefore does not forms a protective layer. This may also explain why $HfO_2$ rather than $Al_2O_3$ can be directly grown on graphene without out-of-plane covalent functional groups, in low-temperature thermal ALD process[41,42].

It is worth emphasizing that high-density 2L graphene islands or uniform distribution Hf-C bonds will favor the coalescence of growing $HfO_2$ islands. Moreover, the height of the 2L graphene islands or the length of Hf-C bonds is in the range of atomic thickness so that the $HfO_2$ thickness could be scaled down to 1-2 nm. It is also worth emphasizing that the presence of the covalent Hf-C bonding is likely to degrade 1L graphene device performance by disrupting the graphene electronic structure. Pirkle et al. fabricated carbide-free $HfO_2$ films on graphene by lowering the deposition temperature to 77 K or promoting the base pressure to $10^{-10}$ mbar in the chamber. However, for 1L-like graphene devices, the Hf-C bonds do not alter the lattice symmetry of the bottom graphene layer. Therefore, 2L graphene may be a prospective material with regard to certain applications. Moreover, it has been reported that low-frequency 1/f noise in the transistor (30-nm gate length) made from 2L graphene is strongly suppressed compared with the monolayer graphene transistors[43].

## 5. Conclusion

We have investigated the structural damage of graphene underlying dielectrics ($HfO_2$ and $Al_2O_3$) by remote plasma-enhanced atomic layer deposition (PE-ALD). We find that the level of damage is decreased with the number of layers. Interestingly, the Raman spectrum of 2L graphene underlying $HfO_2$ resembles that of pristine 1L graphene. XPS measurements indicate the formation of Hf-C carbide. During the initial PE-ALD half-cycle, the upper graphene layer reacts with the TDMA-Hf precursor, creating nucleation islands or a uniform active Hf-C layer. After the monolayer $HfO_2$ coverage, the bottom graphene layer has additional protection. It seems very likely that the upper graphene layer serves as a sacrificial layer, which not only promotes the adhesion of $HfO_2$ on graphene, but also maintains the structural integrity of the bottom graphene layer. Therefore, 2L graphene allows for controlling/limiting the defect formation during the PE-ALD $HfO_2$ process and might be a prospective material for certain applications, such as graphene-based transistors and sensing devices. From the previous results, it appears that the thickness of PE-ALD $HfO_2$ can be arbitrarily scaled down to sub-10-nm range.

Acknowledgements
Xiaohui Tang is a research of FRS-FNRS. The authors thank the staff of UCL's cleanroom facility WINFAB for technical support. This work is financially supported by the Action de Recherche Concertée (ARC) "StressTronics", Communauté française de Belgique. Part of this work is financially supported by the Belgian Fund for Scientific Research (FRS-FNRS) under FRFC contract ''Chemographene'' (No. 2.4577.11). J.-F. Colomer and B. Hackens are Research Associates of FRS-FNRS. This research used resources of the Electron Microscopy Service located at the University of Namur ("Plateforme Technologique Morphologie – Imagerie"). This research used resources of the ELISE Service of the University of Namur. This Service is member of the "Plateforme Technologique SIAM

Figure captions

Figure 1: Scanning electron microscopy images for (a) 1L graphene, (b) 2L graphene, and (c) 3L graphene on copper foils.

Figure 2: Optical microscopy images of (a) pristine graphene on $SiO_2$/Si substrate and (b) HfO2/graphene/ $SiO_2$/Si stack.

Figure 3: Raman spectra of 1L graphene without/with PE-ALD $HfO_2$ and PE-ALD $Al_3O_2$ films for an oxygen plasma power of 300 W.

Figure 4: Raman spectra of 1L graphene covered with (a) PE-ALD $Al_2O_3$ and (b) PE-ALD $HfO_2$ for two different oxygen plasma power levels, respectively, (300 or 150 W) and (300 or 200 W).

Figure 5: Raman spectra of pristine 1L, 2L, and 3L graphene (a). Raman spectra of 1L, 2L, and 3L graphene covered with (b) PE-ALD $Al_2O_3$ and (c) PE-ALD $HfO_2$ at an oxygen plasma power of 150 W and 200 W, respectively.

Figure 6: XPS depth profiles of (a) $Al_2O_3$/silicon, (b) $HfO_2$/silicon, (c) $Al_2O_3$/graphene, and (d) $HfO_2$/graphene samples. C 1$s$ spectra of (e) $Al_2O_3$/graphene and (f) $HfO_2$/graphene samples, corresponding to the maximum of the carbon profiles in (c) and (d), respectively.

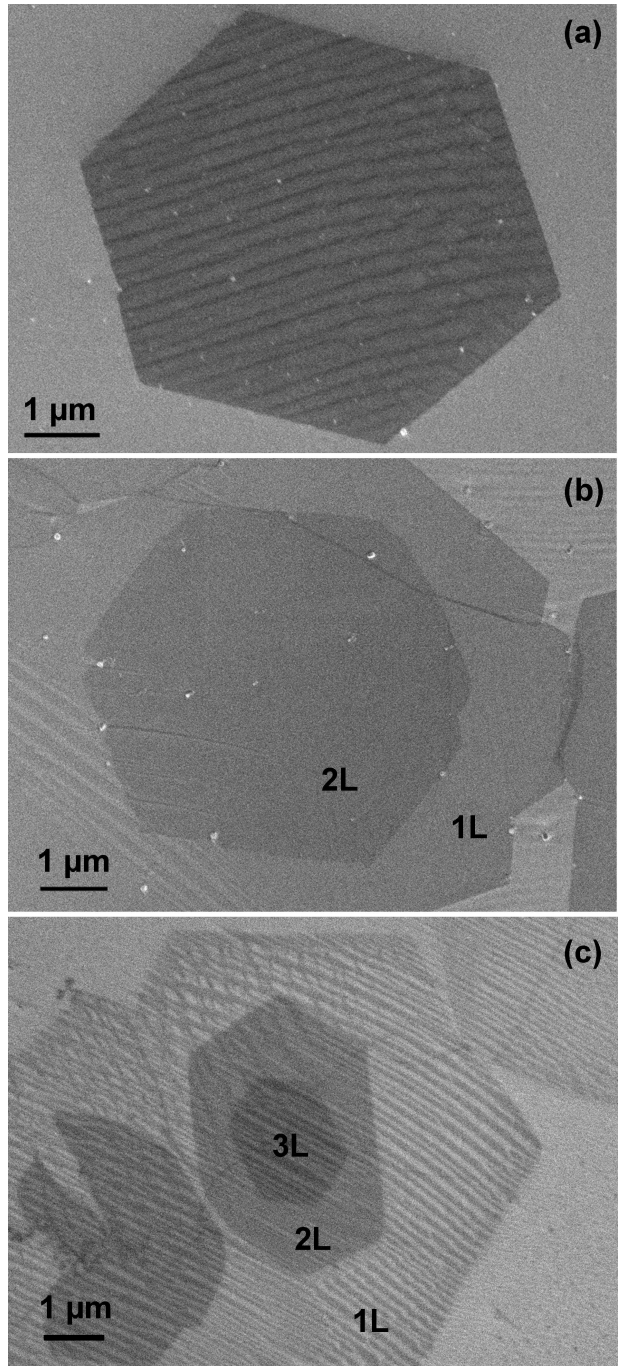

Fig. 1

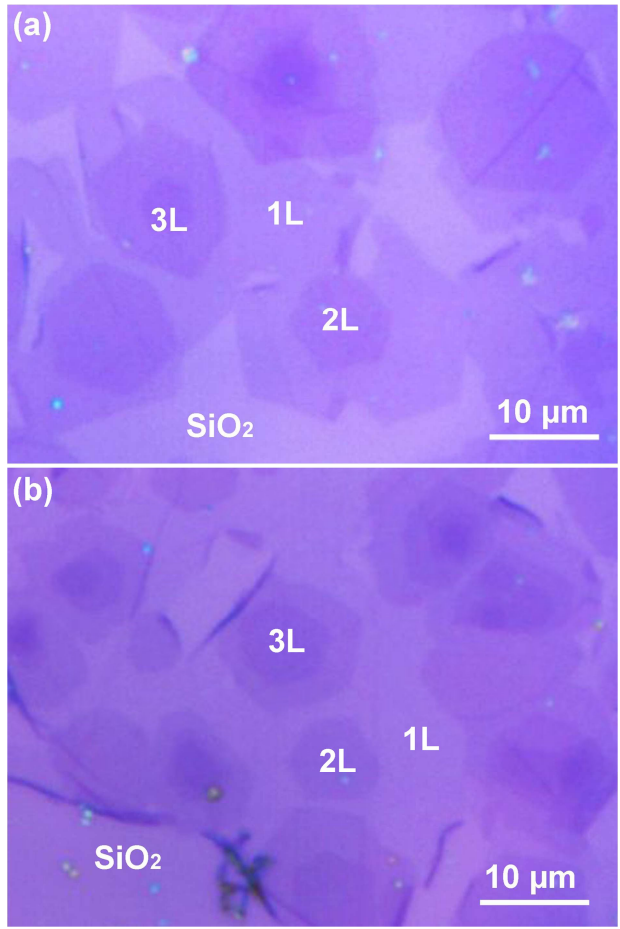

Fig. 2

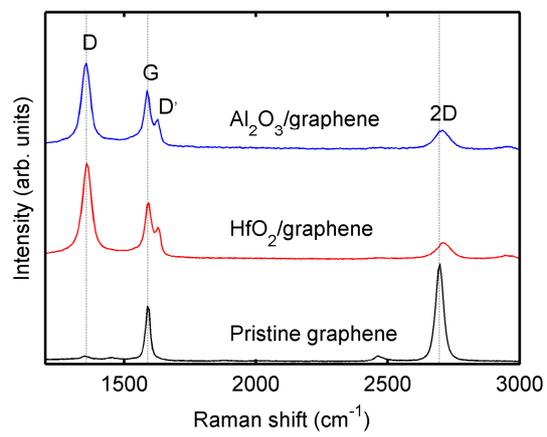

Fig. 3

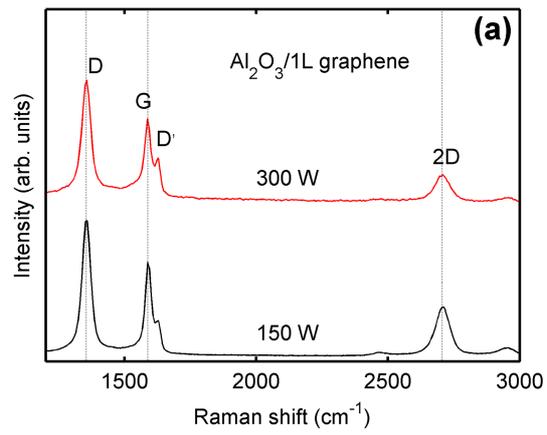

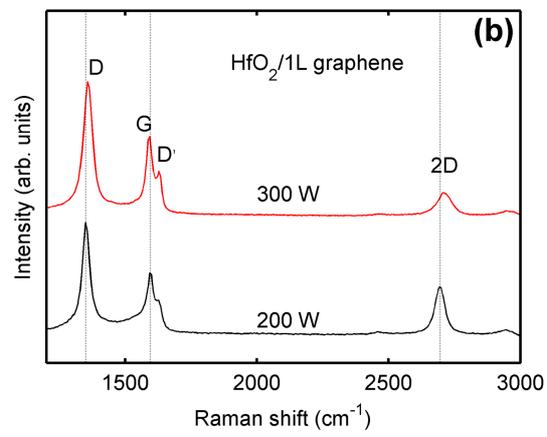

Fig. 4

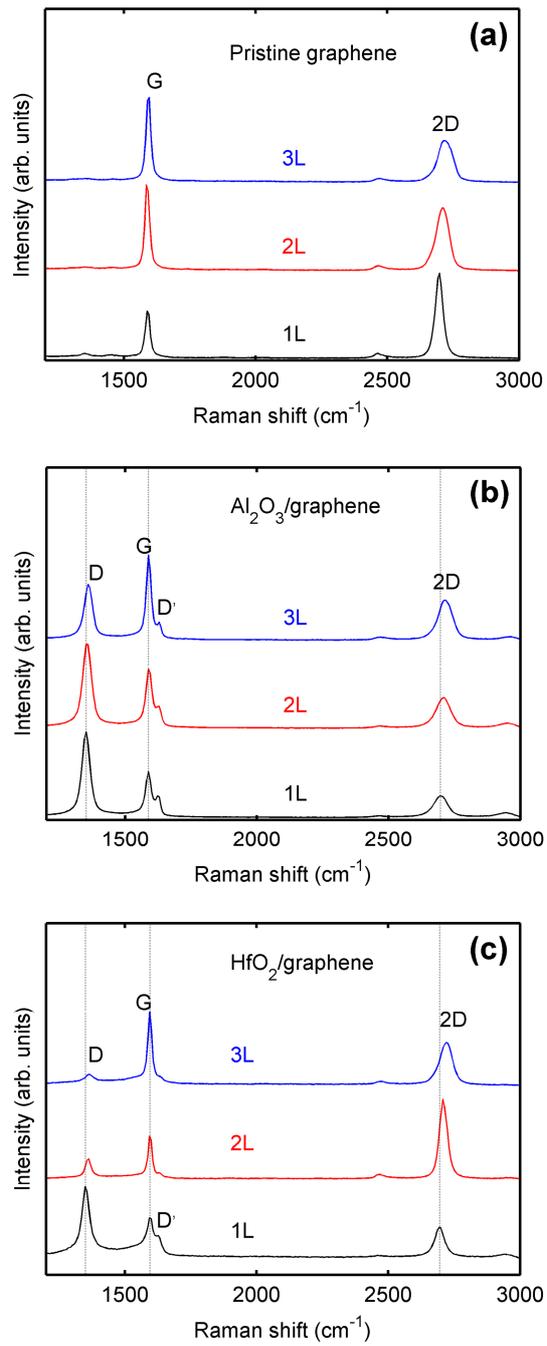

Fig. 5

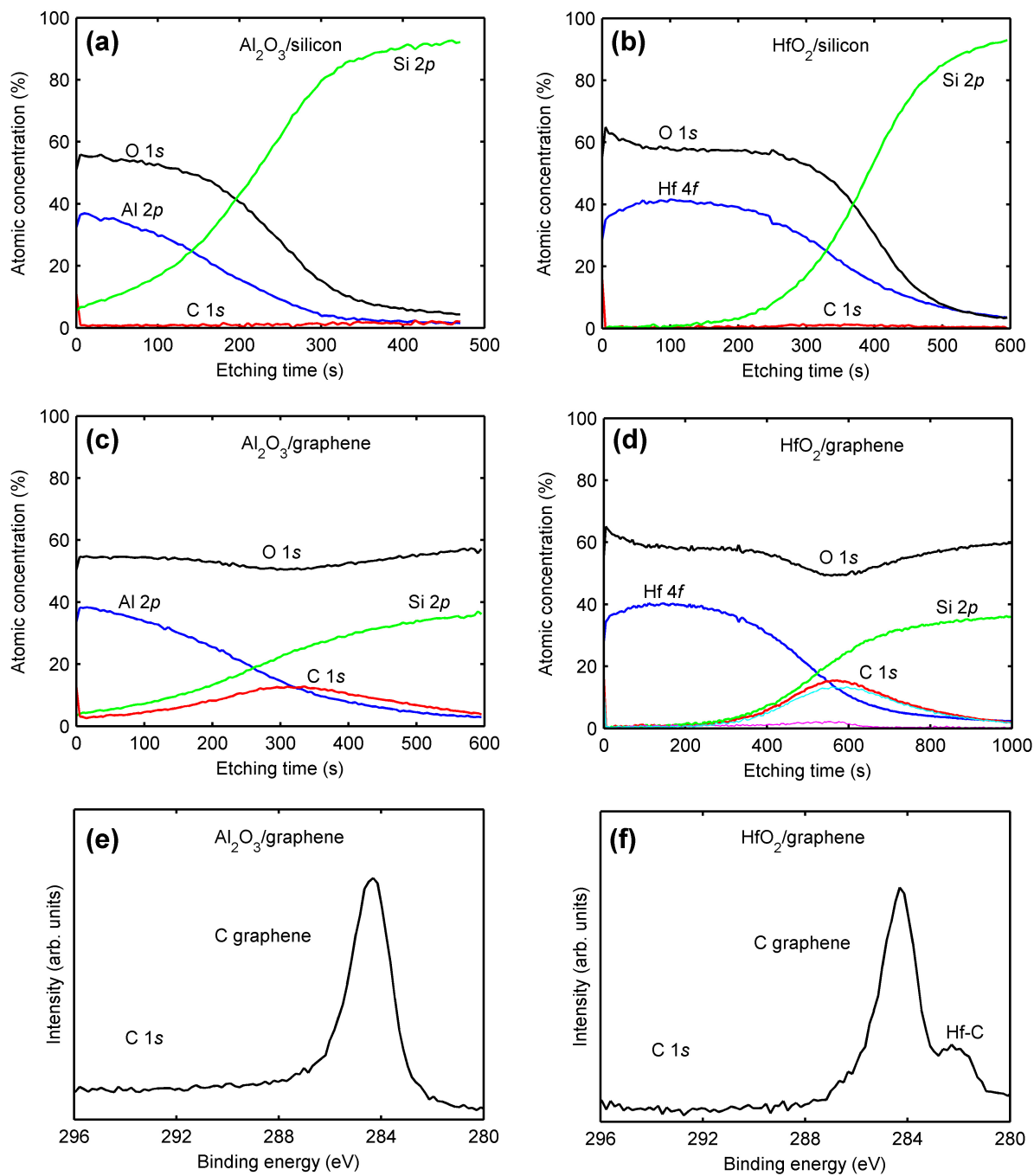

Fig. 6